\documentclass[aps,prb,twocolumn,epsf,floatfix]{revtex4}
\usepackage{txfonts}

\usepackage{epsfig}
\usepackage{color}
\usepackage{graphicx}

\begin{document}

\title{Steady-state solution for dark states using a three-level system in coupled quantum dots}

\author{Tetsufumi Tanamoto}

\affiliation{Corporate R \& D center, Toshiba Corporation,
Saiwai-ku, Kawasaki 212-8582, Japan}

\author{Keiji Ono}
\affiliation{Advanced Science Institute, RIKEN, Wako-shi, Saitama 351-0198,
Japan}

\author{Franco Nori}

\affiliation{The Institute of Physical
and Chemical Research (RIKEN), Wako-shi, Saitama 351-0198,
Japan}
\affiliation{Physics Department, 
The University of Michigan, Ann Arbor, Michigan 48109-1040, USA}

\date{\today}

\begin{abstract}
Quantum dots (QDs) are one of the promising candidates of interconnection between electromagnetic field 
and electrons in solid-state devices. Dark states appear as a result of coherence between the electromagnetic fields 
and the discrete energy levels of the system. Here, we theoretically solve the steady-state solutions of 
the density matrix equations for a thee-level double QD system and investigate the condition 
of the appearance of a dark state. We also numerically show the appearance of the dark state by time-dependent current 
characteristics.
\end{abstract}
\pacs{03.67.Lx, 03.67.Mn, 73.21.La}
\maketitle

\section{Introduction}
Future quantum communication systems might be composed of optical fiber 
networks and quantum computers based on solid-state qubits.
Even though quantum computers made by optical systems would be smoothly connected 
to optical communication networks, in order to interface to mobile electronic systems, 
such as mobile phones, quantum computers based on solid-state circuits are desirable~\cite{You,Rozhkov}. 
In this respect, an efficient interconnection between optical and 
solid-state systems should be developed. 
Quantum dot (QD) systems, such as GaAs/AlGaAs~\cite{Ono,Tarucha,Ono2} have discrete energy levels 
which are suitable for the transfer of photon or phonon energies 
to electrons in solid-state circuits and can be used as elements of 
qubits~\cite{tana0,tanamoto,tana2,Eto,Tokura}. 
QD systems also have the advantage that the distance between energy-levels 
can be controlled by the bias, in addition to the sizes of the QDs. 
Thus QD systems are one of the promising candidates 
for the interconnection between optics and solid-state circuits.

One of the efficient connection methods is constructed by
using coherent population trapping (CPT) or electromagnetically induced transparency (EIT)~\cite{tanaeit}.
CPT is a typical phenomena of quantum coherence in three-level system and 
has been intensively studied in optics~\cite{Harris,Imamoglu,Urabe,Ian,Debald}.
By adjusting two laser fields, the electron population between the lowest two energy-levels are 
coherently transferred. 
Here, we theoretically discuss the interaction between an electromagnetic field 
and a coupled double QD (DQD) system by focusing on transport properties of 
three-level systems.

Recently CPT has been studied in superconducting qubit systems~\cite{Kelly,Chua,Nori}.
Double quantum dot (DQD) systems such as GaAs/AlGaAs\cite{Ono,Tarucha,Ono2} or Si/SiO${}_2$ are also candidates for 
realizing three-level systems and have been theoretically investigated~\cite{Ke,Emary}.
Tokura {\it et al.}~\cite{Tokura} theoretically investigated resonant tunneling currents under locally different Zeeman energies 
and found that 
when the magnetic fields in each QD are non-collinear, four resonant peaks can be observed.
Ke {\it et al.}\cite{Ke} constructed density matrix equations and investigates the relation between 
the phase of the driving lasers and transport properties. Emary {\it et al.}~\cite{Emary} investigated transport properties 
when three-energy levels exist in the same QD. 
From an experimental viewpoint, applying two laser fields is not easy to control.
Here, we mainly discuss the case where one of the laser fields can be replaced by 
electronic tunneling between two QDs.

A three-energy-level DQD system is realized under a large bias voltage as depicted in Fig.1, 
in which there is one energy-level ($E_1$) in the left QD and two ($E_2$ and $E_3$) in the right QD.
We assume a strong Coulomb interaction between electrons such that 
only one excess electron is allowed in the two QDs.
We also assume that the left energy level $E_1$ is close to the right upper energy level 
$E_3$ such that electrons in QD1 tunnel directly into $E_3$ 
($E_3 -E_1 \ll \Omega_L$; and $\Omega_L$ is the tunneling rate between QD1 and QD2).

The dark state is a state in which there is no electron in the $E_3$ level and induces interesting 
phenomena in the transport properties of the DQD system. Let us first think about the case of 
conventional tunneling processes without the dark state: an electron tunnels from the left electrode 
to the QD1 with a tunneling rate $\Gamma_1$. When the laser pulse is switched off, the electron is 
trapped at the $E_2$ level with some probability. Once the electron is trapped at the $E_2$ level, 
because of the Coulomb blockade effect, there is no current through the DQD. 
When the laser pulse is switched on, the electron is excited from the $E_2$ level to the $E_3$ level. 
By the electron tunneling from the $E_3$ level to the right electrode, the current finally flows through the DQD. 
Then, the Coulomb blockade is released, and a new electron can tunnel from the left electrode to the QD1. 
Thus, as long as the laser pulse sequence continues, the current continues to flow.
However, when the dark state is realized, the three-level system is in a coherent superposition state, 
and because there is no electron in the $E_3$ level, the current does not flow through the DQD system 
in spite of the applying laser field and bias voltage.

The purpose of this paper is to show the conditions for the production of a dark state, as functions 
of the Rabi frequency and the detuning parameter of the external laser field.
We derive the density matrix equations in the three-level DQD system, and derive a steady-state solution for a dark state.
We investigate the relationship 
between the time-dependent current characteristic and the dark state.

This article is organized as follows. In \S 2, the formulation 
of our model is presented. In \S 3, we show the analysis of 
the steady-state solution of the dark state.
Section 4 is devoted to the numerical 
calculations for the time-dependent current when there is a dark state. 
The conclusions are given in \S 5.
The detailed analytical solutions of the dark state in its steady state are 
shown in the Appendix. 
We also argue the analysis of the $\rho_{22}=0$ state in the Appendix.

\section{Formulation}
The Hamiltonian is $H=H_0+H_t+H_{l}+H_{\gamma}$, where
\begin{eqnarray}
H_0 &=& E_1 |1\rangle \langle 1|+E_2 |2\rangle \langle 2|
+E_3 |3\rangle \langle 3|, \nonumber \\
H_{t} &=&- ( \Omega_{L} |1\rangle \langle 3|+\Omega_{R}e^{-i\nu_R t}  |2\rangle \langle 3|)
+{\rm h.c.}, 
\nonumber \\
H_{l} &= & \sum_{\alpha\!=L,R} \sum_{k_\alpha}
E_{k_\alpha} |k_\alpha \rangle \langle k_\alpha |,
\nonumber \\
H_{\gamma}&=& \sum_{k_L}V_{L}|k_L \rangle \langle 1|+\sum_{k_R}V_{R}|k_R \rangle \langle 2|+{\rm h.c}. 
\label{eqn:H_set}
\end{eqnarray}

\begin{figure}
\begin{center}
\includegraphics[width=8.5cm]{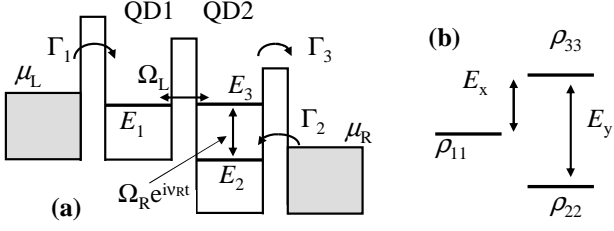}
\end{center}
\vspace*{3cm}
\caption{(a) A three-level system in double QDs~(DQDs). A bias voltage is
applied between the left and right electrodes. (b) A density matrix for the three level. We define 
$E_x\equiv E_3-E_1$ and the detuning $E_y\equiv E_3-E_2-\nu_R$. The dark state is a state with $\rho_{33}=0$.}
\end{figure}
\begin{figure}
\begin{center}
\includegraphics[width=8.5cm]{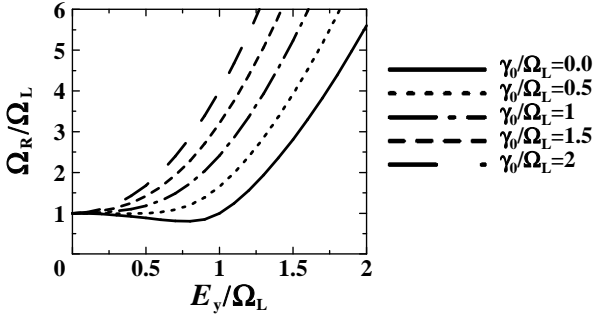}
\vspace*{4cm}
\caption{The Rabi frequency $\Omega_R$ needed for realizing a dark state, as a function of the detuning $E_y$ and $\gamma_0$.
Here, $E_x=0$, $\gamma_{21}=0$, $t_0=0$ and $\Gamma_1=\Gamma_2=\Gamma_3$.
These are solutions of $\rho_{33}=0$ in Eq.(\ref{EqR}).}
\label{Omega_R-Gamma2}
\end{center}
\end{figure}

Here $|i\rangle $~($i=1,2,3$) is the energy-level in QDs,  
$|k_{L}\rangle $~($|k_{R}\rangle $) is the left (right) electrode state. 
$\Omega_R$ is the Rabi frequency between $E_2$ and $E_3$, induced by an external laser field, 
and $\nu_R$ is the laser frequency. 
$V_{L}$~($V_{R}$) are the tunneling 
strengths of electrons between the left (right) 
electrode and the left (right) QD. 
The density matrix $\rho_{ij}\equiv |i\rangle\langle j|$ at $T=0$ is derived 
using
\begin{equation}
\frac{d\rho}{dt}=-\frac{i}{\hbar} [H, \rho ] - \frac{1}{2}\{ \hat{\gamma},\rho\},
\end{equation}
where $\hat{\gamma}$ expresses the dissipation of the system. In the present case of a QD system,
the electron tunnelings between the QD and the electrodes correspond to this dissipation 
(hereafter we set $\hbar=1$).
When we define
\begin{eqnarray}
\tilde{\rho}_{32} &=& \rho_{32} e^{i\nu_R t}, \\
\tilde{\rho}_{21} &=& \rho_{21}  e^{-i\nu_R t},\\
\tilde{\rho}_{31}&=& \rho_{31},  \\
\tilde{\rho}_{ii}&=& \rho_{ii},\ (i=1,2,3) 
\end{eqnarray}
the density matrix equations are given by~\cite{Ke}
\begin{eqnarray}
\frac{d\tilde{\rho}_{00}}{dt} 
\!\!&=&\!\!
  \Gamma^{\rm (h)}_{1}  \tilde{\rho}_{11}
+\Gamma^{\rm (h)}_{2}  \tilde{\rho}_{22} 
+\Gamma^{\rm (h)}_{3}  \tilde{\rho}_{33} 
- \left( \Gamma^{\rm (e)}_{1}  +\Gamma^{\rm (e)}_{2}+\Gamma^{\rm (e)}_{3}\right) \tilde{\rho}_{00},
\nonumber \\
\frac{d\tilde{\rho}_{11}}{dt}\!\!\!&=&\!\!\! 
\Gamma^{\rm (e)}_{1} \tilde{\rho}_{00} - \Gamma^{\rm (h)}_{1} \tilde{\rho}_{11}
+ i\Omega_{L}^* \tilde{\rho}_{31} 
- i\Omega_{L} \tilde{\rho}_{13} 
\nonumber \\
& &+it_0 ( \tilde{\rho}_{12} - \tilde{\rho}_{21}),
\nonumber \\
\frac{d\tilde{\rho}_{22}}{dt} \!\!\!&=&\!\!\!
\Gamma^{\rm (e)}_{2} \tilde{\rho}_{00} - \Gamma^{\rm (h)}_{2} \tilde{\rho}_{22} 
+ i\Omega_{R}^* \tilde{\rho}_{32} 
- i\Omega_{R} \tilde{\rho}_{23} 
\nonumber \\
& &-it_0 ( \tilde{\rho}_{12} - \tilde{\rho}_{21}),
\nonumber \\
\frac{d\tilde{\rho}_{33}}{dt} \!\!\!&=&\!\!\!
\Gamma^{\rm (e)}_{3}  \tilde{\rho}_{00}  -\Gamma^{\rm (h)}_{3} \tilde{\rho}_{33}
+i\Omega_{L}   \tilde{\rho}_{13}
-i\Omega_{L}^*\tilde{\rho}_{31}  
\nonumber \\
& &+i\Omega_{R}    \tilde{\rho}_{23}
-i\Omega_{R}^* \tilde{\rho}_{32},    
\nonumber \\
%
%
\frac{d\tilde{\rho}_{31}}{dt} \!\!\!&=&\!\!\! -(i\omega_{31}+\gamma_{31})\tilde{\rho}_{31}
 -i\Omega_{L} (\tilde{\rho}_{33}-\tilde{\rho}_{11} ) 
+i \Omega_{R} \tilde{\rho}_{21},  \nonumber \\
\frac{d\tilde{\rho}_{32}}{dt} \!\!\!&=&\!\!\! 
-\left[i(\omega_{32}-\nu_R)+\gamma_{32}\right]\tilde{\rho}_{32} 
\nonumber \\
& &-i \Omega_{R} (\tilde{\rho}_{33}-\tilde{\rho}_{22} ) 
+i \Omega_{L} \tilde{\rho}_{12},  
\nonumber \\
%
%
\frac{d\tilde{\rho}_{21}}{dt} \!\!\!&=&\!\!\! 
-\left[i(\omega_{21}+\nu_R)+\gamma_{21}\right]\tilde{\rho}_{21} 
-i\Omega_{L}\tilde{\rho}_{23} +i \Omega_{R}^*  \tilde{\rho}_{31} \nonumber \\
& &-it_0 ( \tilde{\rho}_{11} - \tilde{\rho}_{22}),
\label{dense}
\end{eqnarray}
where  $\gamma_{31}$, $\gamma_{32}$ and $\gamma_{21}$
represent decoherence, such as acoustic phonons. 
In the present conditions, 
$\rho_{i0}$ and $\rho_{0i}$ are given by:
\begin{equation}
\frac{d\tilde{\rho}_{i}}{dt} = 
-i(\omega_{i0}+\gamma_{i0})\tilde{\rho}_{i0}, 
\end{equation}
These equations are solved analytically, but they are independent of the density matrix equations Eqs.~(6), 
therefore, irrelevant to main transport properties;
$\omega_{ij}\equiv E_i-E_j$ ($i=1,2,3)$;
$\Gamma_{i}^{e}$ represents
an electron tunneling from the DQD to the electrodes, and 
$\Gamma_{i}^{(h)}$ represents that from the electrodes
to the DQD, where 
$\Gamma_i^{(x)} \equiv 2\pi \rho_i (E_{Fi})|V_i^{(x)} |^2$, with $\rho_\alpha (E_{Fi})$ 
 ($i=1,2,3$ and $x=e,h$), 
for each electrode at the Fermi energy $E_{Fi}$ ($E_{F1}=\mu_L, E_{F2}=E_{F3}=\mu_R$).

Depending on the relative positions of $E_1$, $E_2$ and $E_3$, 
we can classify the electron transport into the following two regions. \\
(1) $\Gamma_{2}^{(h)}=0$ ($E_2 \gg \mu_R$), \\
(2) $\Gamma_{2}^{(e)}=0$ ($E_2 \ll \mu_R$).\\
Because there is a finite bias between the left and right electrodes, 
the electron does not flow into $E_3$ from the eight electrodes, such that 
$\Gamma_{1}^{(h)}=0$ and $\Gamma_{3}^{(e)}=0$ are satisfied.
In this paper we consider the case shown in Fig.~1 and set 
 $\Gamma_{2}^{(e)}=0$, 
$\Gamma_{1}^{(h)}=0$, and $\Gamma_{3}^{(e)}=0$ 
 (this is the case in which a dark state explicitly exists).
Hereafter we consider $\gamma_{31}=\gamma_0=\gamma_{32}$.

\section{Steady-state solutions for the dark state}
Steady-state solutions are obtained from the density matrix equations 
when $d \rho_{ij}/dt=0$. Compared with the optical three-level~\cite{Brewer}, 
the existence of the $\rho_{00}$ state complicates the equations.
A dark state corresponds to the case where there is no electron state in $E_3$ as $\tilde{\rho}_{33}=0$. 
For a given DQD, we can control the electron tunneling by adjusting the laser field ($\Omega_R$ and $\nu_R$). 
When $t_0=0$, we can express the steady-state solution by fourth-order polynomial 
equations of $E_x$, $E_y$ and $\Omega_R$, such as
\begin{equation}
\tilde{\rho}_{33}  \propto A_{33} (\Omega_R/\Omega_L)^4
+  B_{33} (\Omega_R/\Omega_L)^2 + C_{33} =0,
\label{EqR}
\end{equation}
where 
\begin{eqnarray}
A_{33}&=& \gamma_0'(-\Gamma_2' + \gamma_{21}'), \\
B_{33}&=& -E_y'D_z'\Gamma_2' \gamma_{21}' 
+ {D_z^2}'  ({\Gamma_2}' + {\gamma_0}'){\gamma_0}'  \nonumber \\
&+& {\gamma_{21}}'{\gamma_0}' + {\gamma_{21}^2}'{\gamma_0^2}',
\\
C_{33}&=&  {\Gamma_2}'{\gamma_0}' [ (1 + D_z'E_y')^2 + {\gamma_{21}^2}' {E_y^2}' 
+ ({D_z^2}' + {\gamma_{21}^2}') {\gamma_0^2}' \nonumber \\ 
&+& 2 \gamma_{21}' \gamma_0' ], 
\label{ABC}
\end{eqnarray}
with $D_z\equiv E_x-E_y=E_2-E_1+\nu_R$
(all quantities are rescaled by $\Omega_L$ and indicated by the prime symbol, such as 
$\Gamma_2'=\Gamma_2/\Omega_L$).
This equation is a parabolic function regarding $\Omega_R^2$ with $C_{33}>0$.
Thus,  if $\Gamma_2 > \gamma_{21}$,  $\rho_{33}(\Omega_R^2)=0$ has a solution for positive $\Omega_R^2$. 
Also when $\Gamma_2 > \gamma_{21}$, the coefficient of $\Omega_R^4$ has a negative value, 
therefore, the $\Omega_R^2$ of Eq.(\ref{EqR}) for the dark state is a maximum value for the solutions of the density matrix equations 
to be valid.
Figure~\ref{Omega_R-Gamma2} plots $\Omega_R$, which satisfies  $\rho_{33}(\Omega_R^2)=0$  as a function of $E_y/\Omega_L$.
It can be seen that the larger $\Omega_R$ is required as $E_y$ or decoherence $\gamma_0$ increases.

\begin{figure}
\begin{center}
\includegraphics[width=6.5cm]{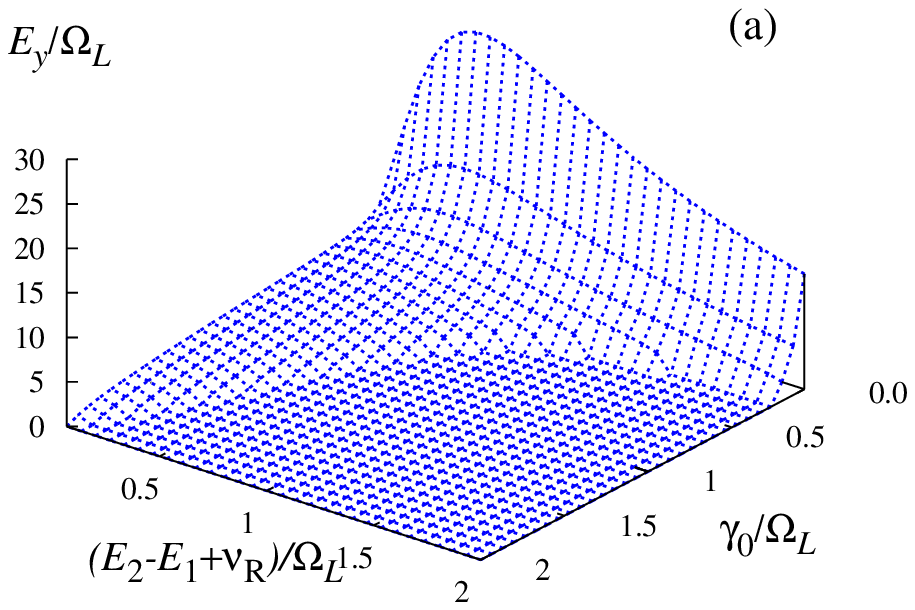}
\\
\vspace*{4.5cm}
\includegraphics[width=6.5cm]{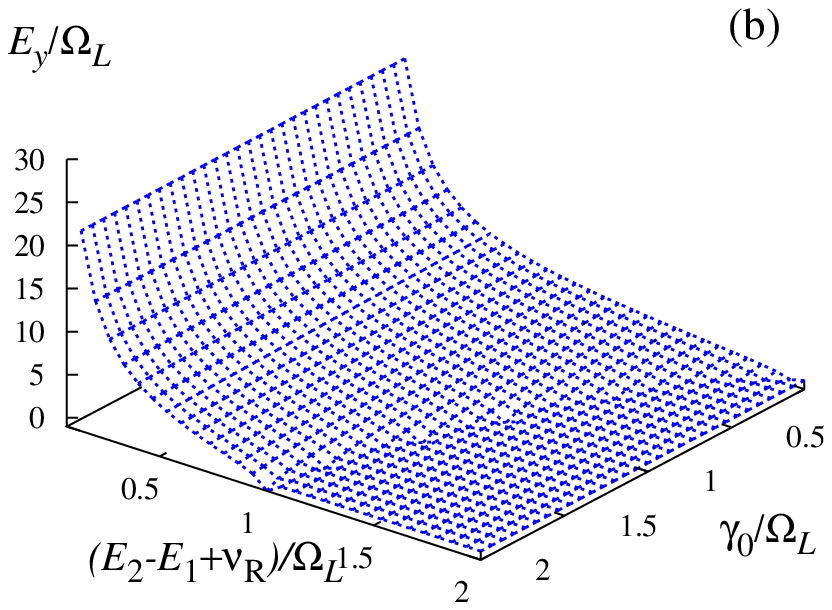}
\end{center}
\vspace*{4.5cm}
\caption{Condition of the dark state of $E_y$ as functions of $E_2-E_1+\nu_R$ and $\gamma_0$ 
when (a) $\gamma_{21}=0.5\Omega_L$ and (b) $\gamma_{21}=0$. Eq.(\ref{EqR}) is solved 
for $E_y$.  We take $t_0=0$, $\Omega_R=2\Omega_L$ and $\Gamma_2=\Omega_L$.}
\label{EY}
\end{figure}

Equation (\ref{ABC}) shows the relationship between $E_y$ and $E_2-E_1+\nu_R$. Figure~\ref{EY} show 
$E_y$ as functions of $E_2-E_1+\nu_R$ and $\gamma_0$ for $\gamma_{21}=0$, and 
$\gamma_{21}=0.5\Omega_L$. We can see that the region of the existence of 
$E_y$ for the dark state becomes smaller in particular when $\gamma_{21}$ becomes larger.
When $\gamma_{21}=0$, $E_y$ is given by
\begin{equation}
 E_y= \frac{1}{D_x\sqrt{\Gamma_2}} \left(-\sqrt{\Gamma_2} + \sqrt{ \Gamma_2\Omega_R^4 
- D_x^2(\Gamma_2 + \gamma_0)\Omega_R^2 - D_x^2\Gamma_2\gamma_0^2  }\right)
\end{equation}
Because the equation in the root square in this equation should have a real solution, we have the
condition:
\begin{equation}
E_2-E_1+\nu_R > \frac{2\Gamma_2\gamma_0}{2\Gamma_2+\gamma_0}.
\end{equation}
Thus, when the decoherence $\gamma_0$ is larger than the tunneling rate $\Gamma$, 
$E_2-E_1+\nu_R >2\Gamma$, and when the decoherence $\gamma_0$ is smaller than 
the tunneling rate $\Gamma$, we have $E_2-E_1+\nu_R >\gamma_0$.

\section{Time-dependent current}
\begin{figure}
\begin{center}
\includegraphics[width=6.5cm]{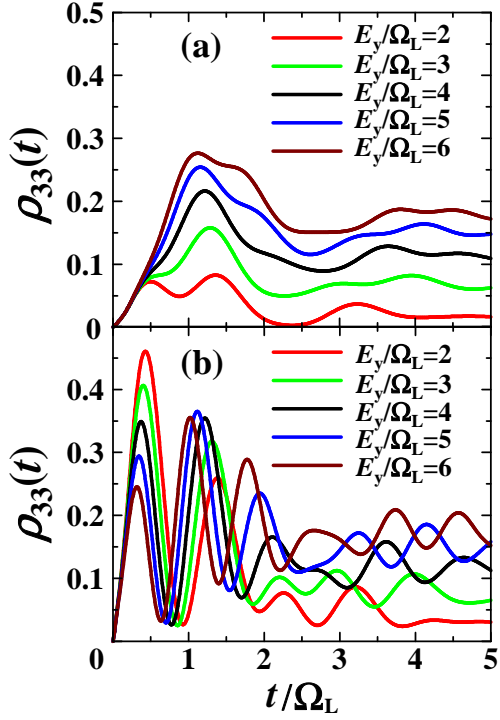}
\end{center}
\vspace*{9.5cm}
\caption{Time-dependent density matrix elements $\rho_{33}(t)$ starting 
from an initial state of (a) $|1\rangle$ and (b) $(|1\rangle+|2\rangle)/2$.
Here, $\gamma_0/\Omega_L=1$, $t_0=0$ and $\gamma_{21}=0$. 
Also, $E_y/\Omega_L=5$ corresponds to the solution of Eq.(\ref{EqR}).}
\label{U33}
\end{figure}

Here we show numerical results for the time-dependent matrix element $\rho_{33}$ and current.
$\Omega_R$ is calculated from Eq.~(\ref{EqR})
such that the initial $E_y$ is given, {\it e.g.}, as $E_y/\Omega_L=2$.
Figure~\ref{U33} shows the time-dependent density matrix element $\rho_{33}$ when 
$E_x=0$, $t_0=0$, and $\gamma_{21}=0$, starting from (a) $|1\rangle$ and (b) $(|1\rangle +|2\rangle)/2$.
It can be seen that, as $E_y$ decreases, $\rho_{33}(t)$ decreases.

As mentioned above, $\Omega_R$ is determined such that it satisfies the steady-state solution 
$\rho_{33} (t\rightarrow \infty) \rightarrow 0$ for $E_y/\Omega_L=2$, where $\rho_{33}(t\rightarrow \infty) $
has the lowest values. Compared with Fig.~\ref{U33}(a), Figure~\ref{U33}(b) oscillates faster. 
This is because, for the superposition state, the density population of electrons oscillates between $|1\rangle$ and 
$|2\rangle$ more often than the case starting from $|1\rangle$.

\begin{figure}[h]
\begin{center}
\includegraphics[width=9cm]{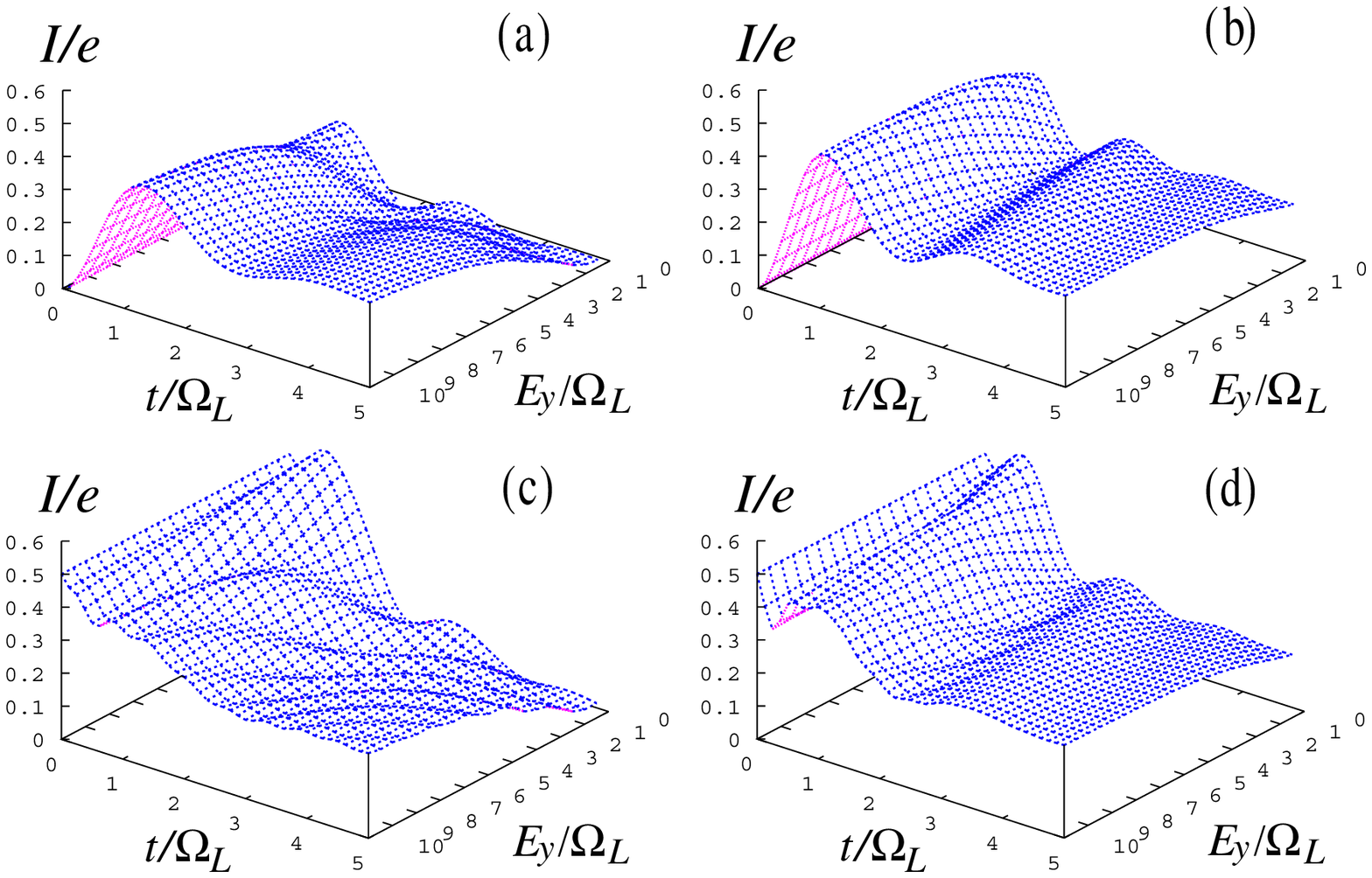}
\end{center}
\label{dme2}
\caption{Time-dependent current as a function of $E_y$ starting from 
$|1\rangle$ for (a) and (b),  $(|1\rangle+|2\rangle)/2$ for (c) and (d).
$\Gamma_1/\Omega_L=\Gamma_2/\Omega_L=\Gamma_3/\Omega_L=1$, 
$\gamma_0/\Omega_L=1$, 
(a)(c) $t_0/\Omega_L=0$ and $\gamma_{21}/\Omega_L=0$.
(b)(d) $t_0/\Omega_L=1$ and $\gamma_{21}/\Omega_L=1$.
}
\label{I1}
\end{figure}

Figure~\ref{I1} shows the time-dependent currents through the DQD system.
The current is derived~\cite{TanaHu} as
\begin{equation}
I(t)=\Gamma_R\left[\rho_{33}(t)+\rho_{22}(t)\right].
\end{equation}
Here we consider 
$\tilde{I}(t)\equiv e^{i\nu_R t} I(t)$.
$\Omega_R$ is determined similarly to Fig~\ref{U33}.
Thus the current is expected to be reduced for $E_y/\Omega_L=2$.
Figures~\ref{I1}(a,c) show that the current decreases  around the expected dark state.
Figures~\ref{I1}(c, d) show the time-dependent currents starting from 
a superposition state of $(|1\rangle + |2\rangle )/2$. 
Compared with Figs.~\ref{I1}(a, c), Figs~\ref{I1}(b, d) show that 
a finite leak tunneling ($t_0=0.5$ and $\gamma_{21}=0.5$) leads to a small current reduction, 
and the evidence of the dark state disappears regardless of the initial state.

\section{Conclusions}
We theoretically solved the steady-state solutions of  the density matrix equations for a thee-level DQD system, 
and showed the condition for the appearance of a dark state.
Numerical calculations for time-dependent current characteristics showed that the
steady-state can be detected by measuring a current.

\acknowledgements
FN was partially supported by
LPS, NSA, ARO, NSF grant No. 0726909, 
JSPS-RFBR contract No. 09-02-92114,
Grant-in-Aid for Scientific Research (S), 
MEXT Kakenhi on Quantum Cybernetics, 
and the JSPS via its FIRST program.
TT thanks A. Nishiyama, J. Koga and S. Fujita for useful discussions.

\appendix
%
\section{Explicit form of density matrix}
Because the equations for $t_0=0$ are still too complicated, we describe the case of $E_x=0$. 
\begin{eqnarray}
\rho_{11}&=& \frac{\Gamma'}{Z_0}
\biggl\{\gamma_0' {\Omega_R^2}' \Bigl(   ({E_y^2}'+ {\gamma_{21}^2}' )\gamma_0' (1 + \Gamma'\gamma_0') 
\nonumber \\
&+& \Gamma' {\Omega_R^2}'(1 + {\Omega_R^2}') + \gamma_{21}'[1 + {\Omega_R^2}' + \Gamma'\gamma_0'(1 + 2{\Omega_R^2}') ]\Bigr) 
\nonumber \\
&+& \Gamma_2'\Bigl(\Gamma'({E_y^2}' + {\gamma_{21}^2}'){\gamma_0^4}' 
\nonumber \\
&+& [{E_y^2}' + \gamma_{21}'(2\Gamma' + \gamma_{21}')]{\gamma_0^3}' (1 + {\Omega_R^2}')
\nonumber \\
&+& {E_y^2}'{\Omega_R^2}'(\gamma_{21}' + \Gamma'{\Omega_R^2}') + \gamma_0'[ {E_y^4}' + (-1 + {\Omega_R^2}')^2 (1 + {\Omega_R^2}')
\nonumber \\
& +& {E_y^2}'(-2 + {\gamma_{21}^2}' + 2{\Omega_R^2}' + 2\Gamma'\gamma_{21}'{\Omega_R^2}')] 
\nonumber \\
&+ & {\gamma_0^2}'[2\gamma_{21}'(1 + {\Omega_R^4}') + \Gamma' h_2]\Bigr) \biggr\},
\\
\rho_{22}&=&
\frac{\Gamma'\gamma_0'{\Omega_R^2}'}{Z_0}
\left[h_1 +  \Gamma'(1 - {E_y^2}' + \gamma_{21}'\gamma_0' + {\Omega_R^2}')\right],
\\
\rho_{33}&=&
\frac{\Gamma'}{Z_0}
\biggl\{\gamma_0'{\Omega_R^2}'h_1 + 
   \Gamma_2' \Bigl( 2\gamma_{21}'{\gamma_0^2}' + ({E_y^2}' + {\gamma_{21}^2}' ){\gamma_0^3}' 
\nonumber \\
&+& {E_y^2}'\gamma_{21}'{\Omega_R^2}' + \gamma_0' [1 + {E_y^4}' - {\Omega_R^4}' 
\nonumber \\
&+& {E_y^2}'(-2 + {\gamma_{21}^2}' + {\Omega_R^2}' )]\Bigr)\biggr\},
\\
Z_0&=& 
  {\Gamma^2}' \Bigl[ {\gamma_0}'{\Omega_R^2}' \bigl( {E_y^2}'(-1 + {\gamma_0^2}' ) + (1 + {\gamma_{21}}'{\gamma_0}' + {\Omega_R^2}' )^2)
\nonumber \\
&+& {\Gamma_2}'\gamma_0'{\Omega_R^2}'h_1 + 
\Gamma_2' \Bigl[( {E_y^2}' + {\gamma_{21}^2}'){\gamma_0^4}' 
\nonumber \\
&+& 2{E_y^2}'\gamma_{21}'\gamma_0'{\Omega_R^2}'
\nonumber \\
&+& {E_y^2}'{\Omega_R^4}' + 2\gamma_{21}'{\gamma_0^3}'(1 + {\Omega_R^2}') 
+ {\gamma_0^2}' h_2 \bigr) \Bigr],
\nonumber \\
&+&  \Gamma' \Bigl[ 4\gamma_0'{\Omega_R^2}'  h_1 + 
    \Gamma_2'\Bigl( 3{E_y^2}'\gamma_{21}'{\Omega_R^2}' 
\nonumber \\
&+& {\gamma_0^3}' ({E_y^2}' + {\gamma_{21}^2}') (3 + {\Omega_R^2}' )
\nonumber \\
&+&  
      \gamma_{21}'{\gamma_0^2}' (6 + {\Omega_R^2}' + 2{\Omega_R^4}' ) 
+ \gamma_0'\bigl(3 + 3{E_y^4}' - 2{\Omega_R^4}' 
\nonumber \\
&+& {\Omega_R^6}' + 
        3{E_y^2}'(-2 + {\gamma_{21}^2}' + {\Omega_R^2}' ) \bigr) \Bigr) \Bigr], 
\end{eqnarray}
where 
\begin{eqnarray}
h_1&=&  ({E_y^2}' + {\gamma_{21}^2}')\gamma_0' + \gamma_{21}'(1 + {\Omega_R^2}'),
 \\
h_2&=& {E_y^4}' + {E_y^2}'(-2 + {\gamma_{21}^2}') + (1 + {\Omega_R^2}')^2.
\end{eqnarray}
Here, all quantities are scaled by $\Omega_L$ such as $\Omega_R'=\Omega_R/\Omega_L$, $\Gamma'=\Gamma/\Omega_L$ and so on.

\section{$\rho_{22}=0$ state}
In the main text, we discussed the dark state condition of $\rho_{33}=0$. 
Here, we consider the region of $\rho_{22}=0$ without considering the dark state. 
Because there is no direct tunneling term between the $E_1$ level and the $E_2$ level, 
an electron exists at the energy-level $E_2$ only when there is some relaxation process of the electron from the $E_3$ level or dissipation.
If the electron is transferred from the left electrode to the right electrode without 
the $E_2$ level, we can regard the system as a two-level system as if  
each QD has one energy-level. 
The condition $\rho_{22}=0$ corresponds to:
\begin{eqnarray}
\rho_{22} &\propto & 
-\gamma_0(\Gamma-\gamma_0 ) (E_x^2+E_y^2) 
\nonumber \\
&+& \left[2\gamma_0( \Gamma-\gamma_0 )+ \Gamma\gamma_{21}\right]E_xE_y
\nonumber \\
&+&  \gamma_0 (\gamma_{21}+ \Gamma) \left[ \Omega_L^2+ \gamma_{21}\gamma_0
+\Omega_R^2\right] =0
\end{eqnarray}
Thus, $E_y$ is written as function of $E_x$:
\begin{equation}
E_y=\frac{1}{a_2}
\left\{ b_2 E_x + \sqrt{(b_2^2-4a_2^2)E_x^2 +4a_2c_2} \right\}
\end{equation}
where
\begin{eqnarray}
a_2&=&\gamma_0 (\Gamma -\gamma_0) \\
b_2&=& a_2 +\Gamma \gamma_{21} \\
c_2&=& \gamma_0 (\gamma_{21}+\Gamma)[\Omega_L^2+\gamma_{21}\gamma_0 +\Omega_R^2 ]. 
\end{eqnarray}
This is the condition that the present system can be treated as a two-level system.
When $\gamma_{21}=0$, we can simplify the condition as:
\begin{equation}
E_y=E_x+\sqrt{ \frac{\Gamma(\Omega_L^2+\Omega_R^2)}{\Gamma -\gamma_0} }.
\end{equation}

\end{document}